\documentclass[aps,prl,twocolumn,floatfix]{revtex4}
\usepackage{graphics,graphicx}
\usepackage{amsfonts}
\usepackage{bm,bbm}
\usepackage{mathrsfs}
\usepackage{calc}
\usepackage{epsfig}
\usepackage{float}
\usepackage{color}
\usepackage{epstopdf}

\def\etal{{\it et al.}}
\def \w{{\omega}}
\def \be{\begin{equation}}
\def \ee{\end{equation}}
\def \ba{\begin{array}}
\def \ea{\end{array}}

\def \partialsin#1{{\frac{d}{d#1}}}
\def \partialbis#1{{\frac{d^2}{d#1^2}}}

\def\bpmat{\begin{pmatrix}}
\def\epmat{\end{pmatrix}}
\def\bmat{\begin{matrix}}
\def\emat{\end{matrix}}

\def\1{\mbox{1\hskip-.25em l}}

\def\beq{\begin{eqnarray}}
\def\eeq{\end{eqnarray}}
\def\be{\begin{equation}}
\def\ee{\end{equation}}
\def\etal{{\it et al. }}
\def\beq{\begin{equation}}
\def\eeq{\end{equation}}

\def \w{{\omega}}
\def \av#1{{\langle#1\rangle}}

\begin{document}
\title{A third-order exceptional point effect on the dynamics of  a single particle in a  time-dependent harmonic trap}
\author{Raam Uzdin$^{(1)}$\footnote{Present address: Department of Physics, The Hebrew University, Jerusalem, Israel.}, Emanuele Dalla Torre$^{(1,2)}$, Ronnie Kosloff$^{(1)}$\footnote{Permanent address: Fritz Haber Institute of Molecular Dynamics, The Hebrew University, Jerusalem, Israel.},  and Nimrod Moiseyev$^{(1)}$\footnote{Permanent address: Schulich Faculty of Chemistry and Faculty of Physics, Technion - Israel Institute of Technology, Haifa 32000 Israel.}}
\affiliation{$^{(1)}$Institute for Theoretical Atomic, Molecular and Optical Physics, Harvard-Smithsonian Center for
Astrophysics and $^{(2)}$Department of Physics, Harvard University, Cambridge, MA 01238, USA}
\date{\today}
\begin{abstract}
The time evolution of a single particle in a harmonic trap with time dependent frequency $\omega(t)$ is well studied. Nevertheless here we show that, when the harmonic trap is opened (or closed) as function of time while keeping the adiabatic parameter $\mu=[d{\omega(t)}/dt]/\omega^2(t)$  fixed, a sharp transition from an oscillatory to a monotonic exponential dynamics occurs at $\mu=2$. At this transition point the time evolution has a third-order exceptional point (EP) \emph{at all instants}. This situation, where an EP of a time-dependent Hermitian Hamiltonian is obtained at any given time, is very different from other known cases. Our finding is relevant  to the dynamics of a single ion in a magnetic, optical, or {\it rf} trap, and of diluted gases of ultracold atoms in  optical traps.
\end{abstract}
\maketitle

\par


Exceptional points (EP) are degeneracies of non-Hermitian Hamiltonians\cite{KATO,Heiss},
associated with the coalescence of two or more eigenstates.
The studies of EPs have substantially grown since the pioneering works of Carl Bender and his co-workers
on ${\cal PT}$-symmetric Hamiltonians\cite{BENDER}. These Hamiltonians have a real spectrum, which becomes
complex at the EP. However, ${\cal PT}$-symmetry is not required to obtain an EP point, as in the case of
a coalescence between two resonant states, leading to
self-orthogonal states\cite{nimrod-friedland,Serra-Edi-nimrod,NHQM-book}.

The physical effects of second-order EPs have already been demonstrated in different types of experiments. See for example the effect of EPs   on cold atoms experiments\cite{Berry0}, on the  cross sections  of electron scattering from hydrogen molecules\cite{EDI}, and on the linewidth of unstable lasers\cite{Berry1}. More direct realizations of EPs in microwave experiments are given in Ref.\cite{RICHTER1,RICHTER2} and in optical experiments in Ref.\cite{MOTI-nature}. For theoretical studies that are relevant to these experiments see for example\cite{EDI,stefan,SK-UG-NM-PRL,DEMETERY,raam,berry,raam-optics}.
In addition, theoretical studies predict significant effects of second-order EPs on the photoionization of atoms\cite{dorr,holger1,holger2} and the photodissociation of molecules\cite{roland,osman,ido}.

The possibility of higher-order EPs (where more than two eigenstates of the non-Hermitian Hamiltonian coalesce at the EP) has been discussed in the literature for time independent  $\cal PT$  symmetric Hamiltonians (see Ref.\cite{Uwe0,EVA-MARIA} and references therein). The main effect of EPs (of any order) on the dynamics of $\cal {PT}$-symmetric systems is the sudden transition from a real spectrum to a complex energy spectrum associated with gain and loss processes.\cite{SK-UG-NM-PRL}.

All above mentioned studies on the effects of EPs are related to non-Hermitian time-independent Hamiltonians. Note that non-Hermitian Hamiltonians can be obtained from Hermitian Hamiltonians  by imposing outgoing boundary conditions  on the eigenfucntions or including complex absorbing potentials\cite{NHQM-book}. This approach allows the description of  resonance phenomena in systems with finite-lifetime metastable states.

Other studies considered time-periodic Hamiltonians where the EPs are
associated with the quasi-energies of the Floquet operator which can be
represented by a time-independent non-Hermitian matrix (see for example one of the first studies   of EP in atomic physics in Ref.\cite{dorr}). EPs were also studied in   non-periodic systems in the context of Landau-Zener-Majorana transitions, where the EP was obtained only after analytic continuation of the actual Hamiltonian\cite{Raam-LZ}. Finally, time-dependent EPs have been used to control the quantum evolution of non-Hermitian systems\cite{RaamUweNimrod}.

In this paper we  show that an EP for the time evolution of a
Hermitian time dependent Hamiltonian can be obtained. Our method relies
on the re-scaling of the time-axis,
allowing to map the time-dependent problem to an effective time-independent one, with non-unitary evolution.


\textit{The harmonic oscillator system with changing frequency in the Heisenberg picture --}
The 1D harmonic oscillator system with changing frequency is defined as
\beq
\label{eq:HAM}
\hat H=\frac{1}{2m}\hat p^2+\frac{1}{2}m\omega^2(t)\hat x^2,
\eeq
where $m$ is the mass of the particle, and $\hat p$ and $\hat x$ are respectively the momentum and position operators. Despite its simplicity, this time-dependent Hermitian Hamiltonian can associated with a third-order exceptional point.

We study the model (\ref{eq:HAM}) in the framework of Ref.\cite{njp,eplY}, where it is shown that, due to the closed commutation relations between the operators $\hat p^2,   \hat x^2, \hat p \hat x+\hat x\hat p$, the model forms a SU(1,1) algebra. As a basis set for this algebra we choose the Hamiltonian, the Lagrangian and the $x-p$ anti-commutator:
\begin{eqnarray}
\hat O_1&=&\hat H\\
\hat O_2&=&\hat L=\hat H-m\omega^2(t)\hat x^2\\
\hat O_3& = &\hat D = \omega(t)(\hat x\hat p+ \hat p \hat x)/2\end{eqnarray}
Any commutator between operators in the algebra can be expressed as a linear combination of these operators:
\beq
[\hat O_k,\hat O_j]_{j=1,2,3}= \sum_{k=1}^3 C_{jk}^l \hat O_l\,\,
\eeq
where the $C_{jk}^l$ are the structure factors of the SU(1,1) algebra\cite{Gilmore}.  The Heisenberg picture for the dynamics
which is associated with the operators $\hat O_j$ is described as
\beq
\label{mov}
\frac{d{\hat O_j}}{dt}=\frac{i}{\hbar}[{\hat H},{\hat O_j}]+\frac{\partial {\hat O}_j}{\partial t},
\eeq
where $j=1,2,3$.

These equations are explicitly given by
\begin{eqnarray}
\begin{array}{l}
\frac{d}{dt} {\hat H} = \frac{\partial \hat H}{\partial t} =  m {\dot {\omega} }\omega \hat x^2 = \omega \mu (\hat H - \hat L)\\
\frac{d}{dt} \hat L =  \frac{i}{\hbar}[\hat H,\hat L]+\frac{\partial \hat L}{\partial t} = -2 \omega \hat D - \omega \mu (\hat H - \hat L)\\
\frac{d}{dt} \hat D =  \frac{i}{\hbar}[\hat H,\hat D]+\frac{\partial \hat D}{\partial t} = 2 \omega L + \omega \mu \hat D\label{eq:EOM}
\end{array}
\end{eqnarray}
Here we defined the dimensionless ``adiabatic parameter''
\beq
\mu=\left[\frac{1}{\omega^2(t)}\right]\frac{d\omega}{dt}.
\eeq
The equations of motion (\ref{eq:EOM}) conserve the ``Casimir'' operator\cite{Uwe} $\hat C(t)=[\hat
H^2(t)-\hat L^2(t) -\hat D^2(t)]/\omega^2(t)$ by satisfying    ${d{\hat C}}/{dt}=0$.

In what follows we will focus on the specific case of $\mu = {\rm const}$, corresponding to the frequency profile
\beq
\label{OMEGA}
\omega (t)=\frac{\omega(0)}{1-\mu\omega(0)t },
\eeq
In experiments the harmonic trap is varied between two extreme values, $\omega_{open}$ and $\omega_{closed}$. The compression factor is given by $\omega_{closed}/\omega_{open}$. For positive values of the adiabatic parameter $\omega(0)=\omega_{open}$  and $\omega(t_f)=\omega_{closed}$.  For negative values $\omega(0)=\omega_{closed}$ and $\omega(t_f)=\omega_{open}$. In both cases $t_f=|\mu|(\omega_{open}^{-1}-\omega_{closed}^{-1})$.

The parameter $\mu$ sets the degree of adiabaticity of the process. For $\mu\to0$, the dynamics is perfectly adiabatic and the system follows the eigenvalues of the instantaneous Hamiltonian.
In contrast, for $\mu\to\pm\infty$, the change of the Hamiltonian is so fast that the system does not have time to change at all. As we will show, these two limits are separated by an exceptional point.  A similar effect is known to occur in the vicinity of quantum critical points (See for example Ref.\cite{toli}) and is here shown in time-dependent non-periodic harmonic traps.

It is convenient to introduce the dimensionless  time variable $\tau=(1/\mu)\log(\omega(t)/\omega(0))$, satisfying $d\tau=\omega (t)dt$, and rewrite (\ref{eq:EOM}) as
\beq
\label{adi}
i\frac{d \bold{\vec O}(\tau)}{d \tau} =  \cal H_{\rm Heis}  \bold{\vec O}(\tau) \equiv \left( {\it i } \mu{\bf I} + \cal H^{\rm Tr=0}_{\rm Heis}\right) \bold{\vec O}(\tau)\label{eq:EOM2}
\eeq
where $\vec O =\{H,L,D\}$-components,  ${\bf I}$ is the unit matrix and the traceless operator ${\cal H}^{\rm Tr=0}_{\rm Heis}$ is defined by
\beq
{\cal H}^{\rm Tr=0}_{\rm Heis}=i
\left(\begin{array}{ccc}
0  &   -\mu & 0  \\
-\mu  &   0 & -2 \\
0       &  2 & 0
\end{array} \right).\label{eq:Heff}
\eeq
We can further simplify Eq.(\ref{eq:EOM2}) by performing the transformation
\be
\label{scaling}
\bold{\vec O}(\tau(t)) \to \frac{1}{\w(t)} \bold{\vec O}(\tau(t)) \quad{\Rightarrow}\quad \frac{d \bold{\vec O}(\tau)}{d \tau} \to \frac{d \bold{\vec O}(\tau)}{d \tau} - \mu{\bf I}\,.
 \ee

The resulting equation of motion $i d\bold{\vec O}/d\tau =\cal H^{\rm Tr=0}_{\rm Heis}\bold{\vec O}$ is equivalent to a time dependent Schr\"odinger equation with a non-Hermitian time-independent Hamiltonian. The matrix $\cal H^{\rm Tr=0}_{\rm Heis}$ is $\cal {PT}$-symmetric\cite{PT-matrix} and its three eigenvalues
\beq E_0=0 \,; \quad E_\pm=\pm \sqrt{4-\mu^2}\,\label{eigs}
\eeq
are real for $|\mu|\le2$. The corresponding eigenvectors are given by, $v_0=\left(1,0,-\mu/2\right),\;v_\pm=\left(\mu,\pm i\sqrt{4-\mu^2}, -2\right)/\mu$.

In contrast to the Schr\"odinger equation, the population of the eigenvectors in a physical state is not arbitrary, but must satisfy several constraints. For example, for $|\mu|<2$, the eigenvectors $v_+$ and $v_-$ are complex and any physical state must populate them with an equal weight, in order to keep the expectation values $H$, $D$, and $L$ real. In addition, $v_+$ and $v_-$ have a zero Casimir constant $\langle\hat C\rangle = H^{2}-L^{2}-D^{2}=0$. Due to the uncertainty relation $\langle\hat C\rangle \ge \hbar^2/4$ any physical state must necessarily populate the eigenstate $v_0$ with non-zero weight as well. Thus, for a generic initial state, we expect more than one eigenvector to be occupied, leading to an oscillatory behavior that we describe below.

\textit{The third-order EP for the time evolution operator of the Hermitian time-dependent harmonic oscillator --} The matrix $\cal H^{\rm Tr=0}_{\rm hies}$ has a third order EP at $|\mu|=2$. At this point all three eigenvalues and the corresponding eigenvectors (in the $HLD$ space) coalesce. As a consequence, $\Big[{\cal H}^{\rm Tr=0}_{\rm Heis}(\mu=\pm 2)\Big]^3=0\,$ while $\Big[{\cal H}^{\rm Tr=0}_{\rm Heis}(\mu = \pm 2)\Big]^2\ne0$, demonstrating that the present EP is of third order.

At the EP the matrix $\cal H^{\rm Tr=0}_{\rm Heis}$ has one single eigenvector, which is  ``self-orthogonal''. To show this property, it is necessary to multiply the right and left  eigenvectors of the non-symmetric $\cal H^{\rm Tr=0}_{\rm Heis}$. At the EP the right eigenvector is $\left(1,0,-1\right)$ while the left eigenvector is $\left(1,0,1\right)$. Their product is equal to zero showing that the eigenvector is ``self-orthogonal'' (see for example Chapter 9 in Ref.\cite{NHQM-book}).


Let us first discuss the situation where the initial state is
 Gaussian (e.g., a ground state  of the harmonic oscillator trap at $t=0$ or a thermal state).
The evolution in time is described  by a  Gaussian Wigner distribution with elliptic contour plots.
Using the normalized coordinates, $x(m\omega)^{1/2}$ and $p/(m\omega)^{1/2}$, the standard deviation of the narrow axis (N) and the wide axis  (W),
are given by:
\begin{equation}
 \sigma_{W,N}^{2}=\frac{H\pm\sqrt{L^{2}+D^{2}}}{\omega}\,.
\end{equation}
Note that by virtue of the Casimir constant a most general state satisfies $\sigma_N\sigma_W\ge 1/2$.
For $\omega=\mbox{const}$ ($\mu=0$) the evolution
simply mixes $L$ and $D$ leaving $\sqrt{L^{2}+D^{2}}$ constant. Hence
even though for $\omega=\mbox{const}$ the distribution rotates in phase space
and changes the variance of position and momentum, the
width of the narrow and wide axis of the distribution remains fixed.

For $|\mu|<2$  time oscillations of measurable quantities  are obtained.
A convenient measure to capture this oscillatory dynamics
is given by the ratio:
\begin{equation}
\label{ratio}
\rho=\frac{\sigma_{W}}{\sigma_{N}}
\end{equation}
where by definition $\rho\ge 1$.
The time evolution of $\rho$ is shown in Fig.~1a as function of the time-dependent compression factor $\omega(t)/\omega(0)$ in  a log scale.
Since the logarithm of the compression factor is equal to the new time variable $\tau$ (multiplied  by $\mu$), in this scale $\rho(t)$ shows periodic oscillations. In the original time variable $t$,  the system displays non-periodic oscillations  with the same amplitude.
The periodicity in the $\tau$ time coordinate is determined by
the eigenvalue difference:
\begin{equation}
\label{tau}
T_{\tau}=\frac{2\pi}{\frac{1}{2}(E_{+}-E_{-})}=\frac{2\pi}{\sqrt{4-\mu^{2}}}
\end{equation}

The visibility of the  fringes pattern that appear in Fig. 1a
is given by the simple expression:
\begin{equation}
\label{visibility}
V=\frac{\rho_{max}-\rho_{min}}{\rho_{max}+\rho_{min}}=\frac{|\mu|}{2}\,.\label{eq:ratio}
\end{equation}
At $|\mu|=2$ the visibility becomes one, and the oscillations disappear for $|\mu|\ge2$.

We now comment on the time evolution starting from a generic non-Gaussian initial state.
Of course, in this case $H$,$L$ and $D$ are not sufficient to  completely determine
the state. Nevertheless for $\mu=\mbox{const}$ the expectation values of H,L and D follow exactly the dynamics described above. It is possible to show that the expression (\ref{eq:ratio}) is valid for any initial state satisfying $\mu D(0) + L(0)^2/(4 H(0)) > 0$. In particular, this condition is fulfilled by any initial state that is stationary with respect to the initial Hamiltonian $\hat H(0)$.

%

A similar transition between oscillatory and monotonous behavior appears in other properly
scaled quantities (see Eq.\ref{scaling}), like $\langle \hat H-\hat L \rangle/\omega(t) = m\omega(t)\left\langle  x^{2}\right\rangle$ and $\langle \hat H + \hat L \rangle/\omega(t)=\left\langle p^{2}\right\rangle/(m\omega(t))$, as shown in Fig.~1b.
However note that for initial states which are not eigenstates of the harmonic trap these quantities
display   additional  trivial oscillations that show up even for  $\omega=\mbox{const}$.

\textit{The EP  corresponds to a transition from an under-damped to an over-damped harmonic oscillator--}
We now present a different approach which will clarify the relation between the present problem and energy-dissipative systems. Our approach is based on the equivalence between quantum and classical evolution of {\it quadratic} Hamiltonians. To reproduce the quantum mechanical results one simply needs to complement the classical equations of motion by stochastic initial distributions, given by the Wigner transform of the initial state.

In our case, the relevant equation of motion is Newton's law
\be \left[\partialbis{t} + \w^2(t)\right] x(t) = 0 \label{eq:eom}\ee
By applying the transformation $d\tau =\w(t) dt$, or $\partialsin{t} = \w(t)\partialsin{\tau}$, we obtain
\be \partialbis{t} x = \partialsin{t} (\w(t) \partialsin{\tau} x) = \w'(t) \partialsin{\tau} f + \w^2(t) \partialsin{\tau} x \ee
In the specific case $\mu = {\rm const}$, the equation of motion then becomes
\be \left[\partialbis{\tau} + \mu \partialsin{\tau} + 1\right] x(\tau) = 0\label{eq:damped}\ee
 Here we obtain the time-independent non-Hermitian equation of motion of a damped harmonic oscillator. The EP $|\mu|=2$ corresponds to the transition between an under-damped and over-damped oscillator, as can be seen by the Fourier transform of (\ref{eq:damped}), leading to:
\be \lambda_{\pm} = \frac{ i\mu \pm \sqrt{4-\mu^2}}2 \label{eq:lambda} \ee
For $|\mu|>2$ both eigenfrequencies are pure imaginary, leading to the disappearance of the oscillatory behavior.


\begin{figure}[h]
  \includegraphics[height=4cm,width=8cm]{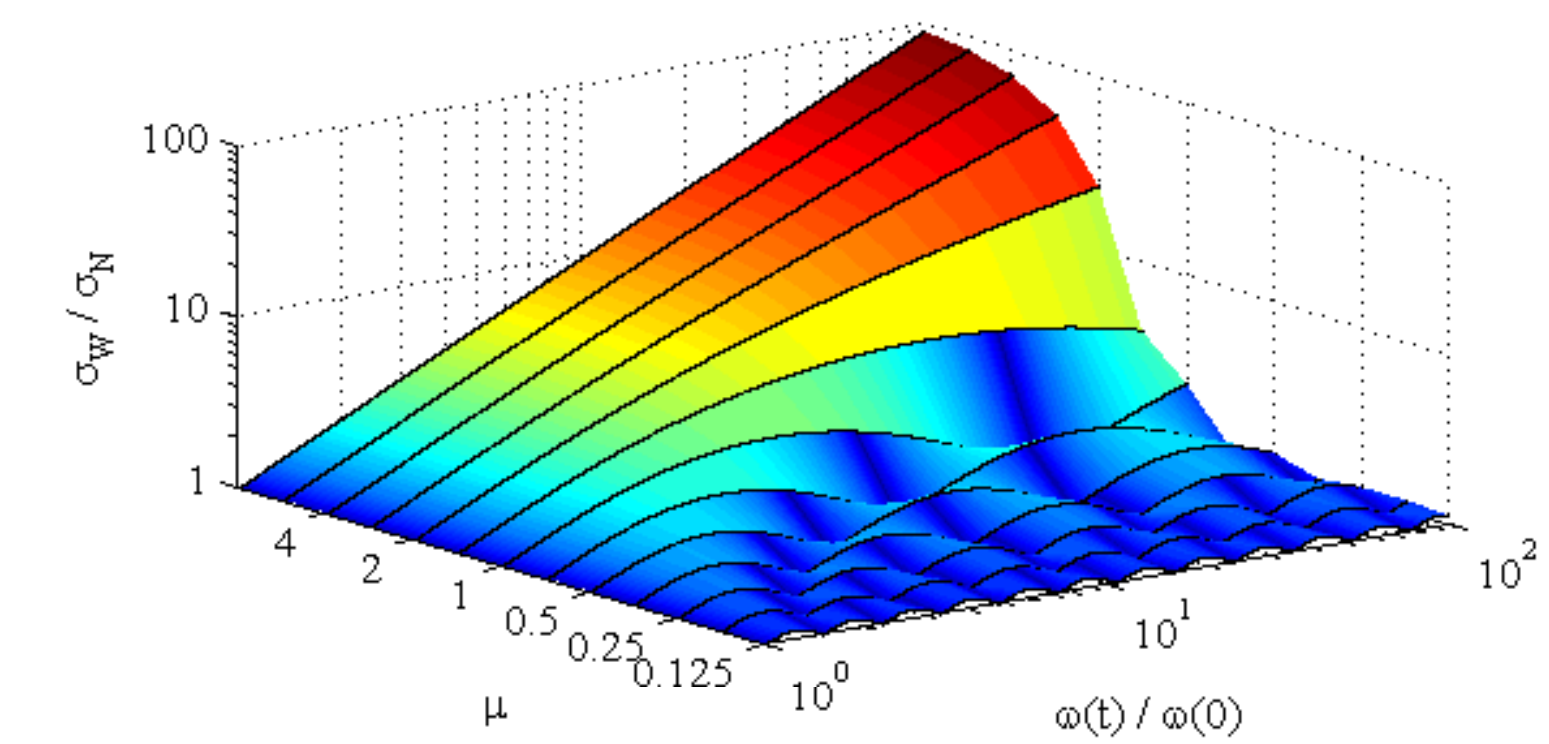}\\
  \centering{(a)}\\
  \includegraphics[height=4cm,width=8cm]{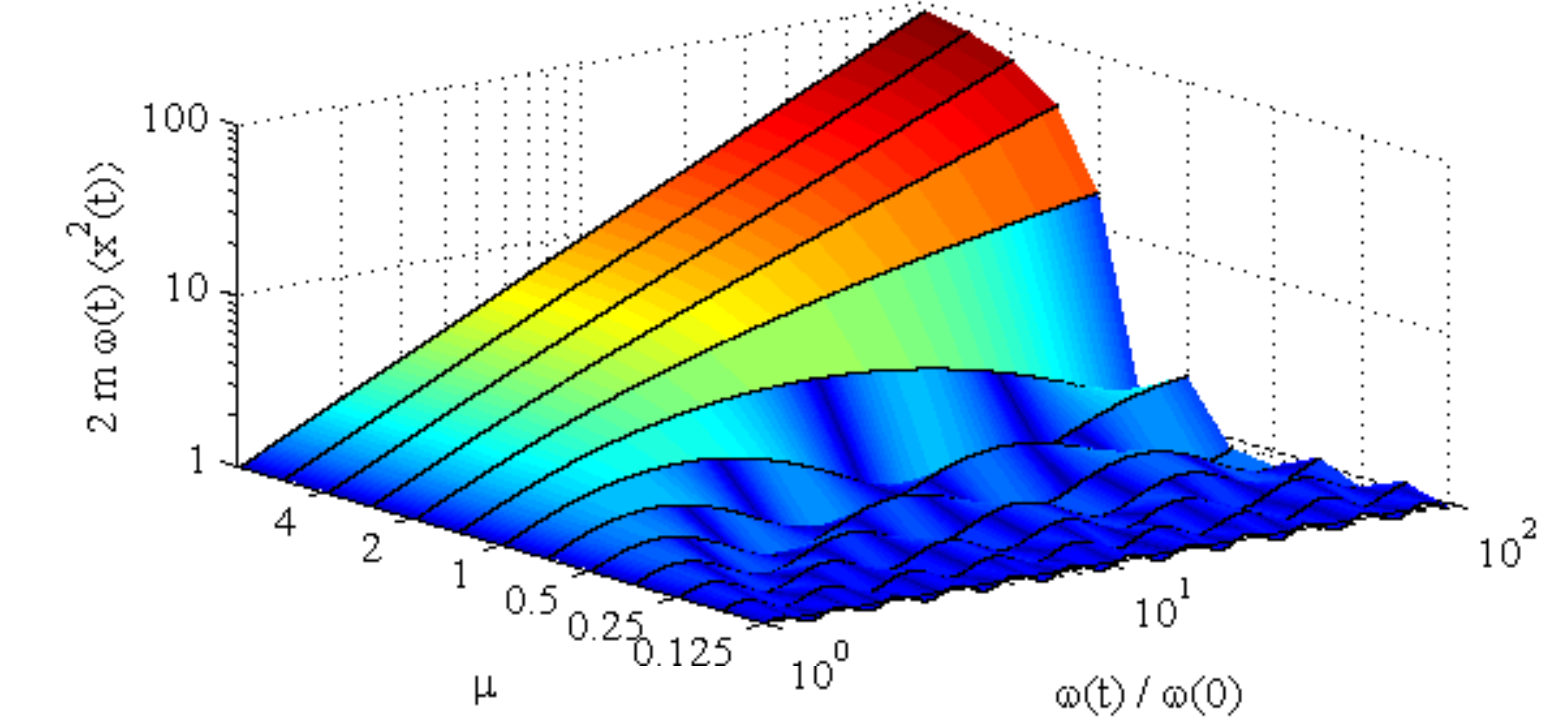}\\
  \centering{(b)}
 \caption{(a) The ratio, $\rho$, between the narrow and wide axis in the Gaussian Wigner  distribution  as a function of the time dependent compression factor $\omega(t)/\omega(0)$ for different values of the adiabatic parameter $\mu\ge 0$. For $\mu<0$ the same plot is obtained where now  the compression factor is taken as $\omega(0)/\omega(t)$. At the exceptional  point (EP) $\mu=2$ the dynamics changes from oscillatory to monotonous. (b) Same plot for the variance of the position operator normalized by the instantaneous frequency, $2\av{x^2(t)}\w(t)$. The initial state is the ground state of the Hamiltonian (\ref{eq:HAM}) at $t=0$ where $\langle x^2\rangle=1/(2m\omega(0))$.}
  \end{figure}



{\it Physical realization of the third-order EP in experiments --} Although our Hamiltonian can be realized in any controllable Harmonics trap (optics, plasma, ...), we will consider here the case of either a single particle (ion\cite{abah}), or a dilute atomic cloud\cite{momentum-distribution1,momentum-distribution2,momentum-distribution3}, in  time-dependent confining traps. The realization with atomic clouds allows the measurement of expectation values in a single-shot experiment. Complications of the dynamics due to the atom-atom interactions  can be avoided (minimized) by setting the atomic scattering length to zero in the vicinity of  a Feshbach resonance\cite{FeshbachRMP}.

The procedure to observe the EP effect on the dynamics of time-dependent Hermitian Hamiltonian  is as follows:\\
(1) Equilibrate the matter in a harmonic trap characterized by the frequency $\omega(0)$. \\
(2) Vary the frequency of the trap as function of time from $\w(0)$ to $\w(t_f)$, while imposing a constant adiabatic parameter $\mu$, as shown in Eq.\ref{OMEGA}.\\
(3) At time $\{0 <t_n<t_f\}_{n=1,2...,N}$ measure either the spatial distribution of matter inside the trap\cite{spatial-distribution1,spatial-distribution2,spatial-distribution3,momentum-distribution3,spatial-distribution4,spatial-distribution5},  or the momentum distribution in a time of flight experiment, by suddenly turning off the trap.\\
(4) Scale the variance of the measured data (position or momentum) by the instantaneous frequency: $\langle{x^2(t)}\rangle \to \omega(t)\langle{x^2(t)}\rangle$ and $\langle{p^2(t)}\rangle \to \langle{x^2(t)}\rangle/\omega(t)$. These quantities are then plotted as function of the compression factor $\omega(t)/\omega(0)$ for different values of $\mu$ as in Fig.1b, showing oscillatory behavior for any $|\mu|<2$, and exponential  behavior for $|\mu|\ge 2$.\\

To avoid the need to use an eigenstate of the Hamiltonian at $t=0$,
one should plot the ratio between the narrow and wide axis of the Wigner distribution of the propagated wavepacket $\sigma_N$ and $\sigma_W$, defined above. This ratio can be measured as follows. At time $t_n$, rather than performing a direct measurement, we propose to keep the frequency of the trap unchanged at $\omega=\omega(t_n)$ and to measure the variance of the position (or momentum) as function of time. The minimum and the maximum of $m\omega(t_n)\langle x^2(t)\rangle$ are respectively $\sigma_N^2$ and $\sigma_W^2$.   This method is perhaps more time-consuming, but guarantees the independence of the result on the initial preparation provided that $\mu D(0) + L(0)^2/(4 H(0)) > 0$.  Note that an initial stationary state (either an eigenstate of the initial Hamiltonian or a thermal state) satisfies this condition.

\textit{Concluding remarks --}
The dramatic effect of EPs  of non-Hermitian time-independent Hamiltonian systems on the dynamics is in the focus of recent theoretical and experimental studies in various fields of physics (as for example in optical or microwave experiments where the material has a complex index of refraction).
Here we  show that the dynamics of a system described by a \emph{time-dependent Hermitian Hamiltonian} can be strongly affected  by a \emph{third-order EP} of an effective  time-independent Hamiltonian.
The fact that the dynamics of the Hermitian time-dependent harmonic oscillator can be explained by the existence of an EP \emph{at all instants} shows the richness of the dynamics of one of the most basic model Hamiltonians, which constitutes a milestone in a large variety of fields in physics. Our finding is both interesting for fundamental theoretical reasons and useful to control the dynamics of a single-ion and of a diluted BEC in a time-dependent traps.

\begin{acknowledgments}
We would like to acknowledge ITAMP for the support.  This work was initialized and has been carried out as  a result from the friendly and scientific atmosphere in ITAMP which encourages informal discussions on a daily basis among the permanent and visiting members of ITAMP.
\end{acknowledgments}

\end{document}